\documentclass[prd,aps,preprint,nofootinbibsuperscriptaddress,tightenlines]{revtex4}
\usepackage{graphicx}
\usepackage{epsfig}
\usepackage{amsmath}
\usepackage{bm}
\begin{document}

\title{Multiple scattering and energy loss 
in semi-inclusive deeply inelastic eA scattering}

\author{Xiaofeng Guo}
\email{gxf@iastate.edu} 
\affiliation{Department of Physics and Astronomy, 
             Iowa State University, Ames, IA 50011}
\author{Jun Li}
\email{junli@iastate.edu} 
\affiliation{Department of Physics and Astronomy, Iowa State
  University, Ames, IA 50011} 
\date{\today}

\begin{abstract}
We calculate the multiple scattering effect 
on single hadron production in semi-inclusive lepton-nucleus 
deeply inelastic scattering. We show that the quantum 
interference of multiple scattering amplitudes leads to suppression 
in hadron productions. At the leading power in medium length, 
the suppression can be approximately expressed in terms
of a shift in $z$ of the fragmentation function $D(z)$, and could
be therefore interpreted as the collisional energy loss. 
We compare our calculation with existing experimental data. 
We also discuss the effect of quark mass on the suppression.
Our approach can be extended to other observables in hadronic 
collisions.
\end{abstract}


\maketitle

\newcommand{\be}{\begin{equation}}
\newcommand{\ee}{\end{equation}}
\newcommand{\ben}{\[}
\newcommand{\een}{\]}
\newcommand{\ba}{\begin{eqnarray}}
\newcommand{\ea}{\end{eqnarray}}
\newcommand{\Tr}{{\rm Tr} }
\section{Introduction}
The understanding of parton propagation through the nuclear environment 
is crucial for the interpretation of physics phenomena observed at 
the Relativistic Heavy Ion Collider(RHIC) and future 
Large hadron Collider(LHC). The observed strong suppression 
of high transverse momentum hadron at RHIC
was considered to be an evidence for the  QCD quark-gluon 
plasma \cite{rhic-pion}.
The suppression was believed to be the result of medium induced
radiative energy loss of high energy partons \cite{jet-quenching}.
However, recent data indicate that heavy quarks would have to lose
the same amount of energy as that of a light quark
if the radiative energy loss is the only source of the suppression
\cite{charm-loss}.  On the other hand, we expect heavy quarks to lose
much less energy than a light quark because of its mass \cite{dead-cone}.
This discrepancy attracted significant theoretical interests in 
searching for other causes of parton energy loss \cite{other-source}.
Several studies suggested that the collisional energy loss
could be an important source of the observed discrepancy \cite{collisional}. 
Because of QCD confinement, we can not observe partons directly
in experiments, instead, we can only observe final state hadrons. 
The production rate could be affected by the interference of
two scattering amplitudes with the same initial and final hadronic 
states but with different partonic interactions.
In this paper, we study the effect of such quantum interference 
between two amplitudes with the same as well as different multiple 
parton-level scattering, and show that the interference leads to
a suppression in single hadron production rate, which might be 
interpreted as the collisional energy loss \cite{other-source}. 

The hadronization of partons is a non-perturbative process.
However, when the energy scale of the scattering $Q$ is much larger
than a typical hadronic scale 1/fm$\sim\Lambda^2_{QCD}$, 
it is the QCD factorization that allows us to separate the calculable 
short-distance partonic dynamics from the non-perturbative long-distance 
physics. The effect of the non-perturbative hadronization process 
for a parton of flavor $f$ fragmenting to a hadron $h$
is expressed in terms of an universal fragmentation function 
$D_{f\rightarrow h}(z,\mu_F)$, where $z$ is momentum fraction of the
parton carried by the hadron and $\mu_F$ is the fragmentation scale. 
QCD perturbation theory predicts the evolution and the scale dependence
on $\mu_F$ of the fragmentation function.

Unlike in the vacuum,  the fragmenting parton in nuclear medium can 
have rescattering before the formation of final-state hadrons.  
The effect of the interaction between the nuclear medium and 
the propagating parton will manifest itself
as changes in parton fragmentation functions.
The rescattering 
can induce extra radiations, and consequently alter the evolution 
of the fragmentation functions \cite{Guo:2000nz, Wang:2001if}, 
and results in effective parton energy loss \cite{Wang-Wang}.

However, quantum mechanically, the parton rescattering in a
nuclear medium does not have to induce radiation.  Such rescattering 
can also alter the production rate of the fragmenting parton if 
it is quantum coherent with the first scattering.  Although
coherent rescattering is formally suppressed by additional powers of
the hard scale, it could be important if the life time of the 
fragmenting parton is long enough.  We show in this paper that
quantum interference of two scattering amplitudes with 
different parton-level rescattering reduces the production rate 
of the fragmenting parton, which leads to the suppression of 
hadron productions.  In the following sections, we show that
such suppression could be effectively expressed as a shift in $z$ 
for the parton fragmentation functions. This result is complementary 
to the radiative parton energy loss induced by the rescattering 
of the fragmenting parton in the nuclear medium.

In general, the production of hadrons can come from both quarks and 
gluons in hadronic production. However, in the semi-inclusive 
deep-inelastic lepton-nucleus scattering (SIDIS), the leading hadron
production is dominated by quarks. It is the ideal place to study the
quark energy loss and the knowledge obtained will help us learn 
more about the gluon energy loss in hadronic productions. 
 
Our paper is organized as follows. In the next section, we first present 
our derivation of double scattering effect. Then in Sec.~III, we show 
how to generalize it to the $n$th-scattering and sum up all possible
numbers of scatterings, and present the result after the summation.
In Sec.~IV, we extend our results for light quarks to quarks 
with finite mass and discuss the effect of quark mass.
In Sec.~V, we compare our results with experimental data from HERMES.
We also discuss the applicable ranges of $z$ value for our result and 
propose a model for larger $z$ region. Finally, in Sec.~VI, we summarize 
our work and discuss extensions of our approach to other observables in
hadronic collisions. 
 
\begin{figure}
\begin{center}
\includegraphics[width=2.5in]{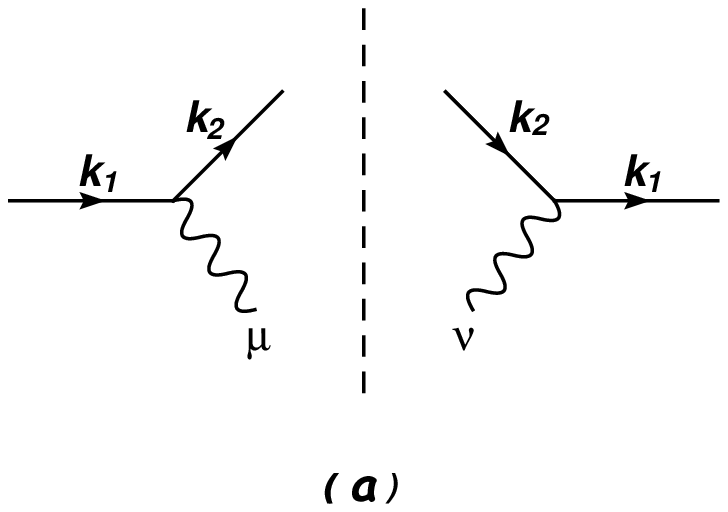}
\hskip 0.5in
\includegraphics[width=2.5in]{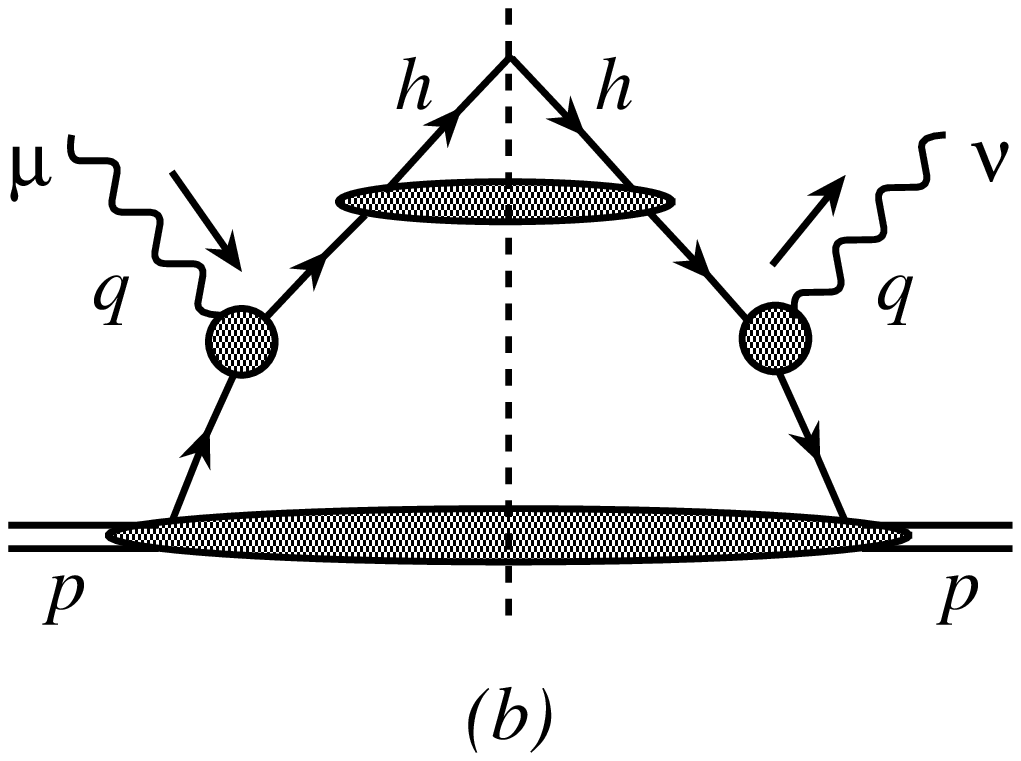}
\caption{Factorization for semi-inclusive DIS cross section:
(a): the leptonic tensor $L^{\mu\nu}$;
(b): the semi-inclusive hadronic tensor.}
\label{fig1}
\end{center}
\end{figure}

\section{Double scattering contribution}

We consider the semi-inclusive DIS production of a single hadron of 
momentum $p_h$,
\begin{equation}
e(k_1) + A(p) \longrightarrow e(k_2) + h(p_h) +X \ ,
\label{process}
\end{equation}
where $k_1$ and $k_2$ are the four
momenta of the incoming and the outgoing leptons respectively,  
$p$ is the average momentum per nucleon for the nucleus with the atomic 
number $A$. 
To study the nuclear effects,  
we compute the hadron production rate per DIS event, 
\begin{equation}
R^A = \left. 
\frac{d\sigma_{eA\rightarrow e h  X}}{dx_B  dQ^2 dz} 
\right/ \frac{d\sigma_{eA\rightarrow eX}}{dx_B  dQ^2} \, ;
\label{rate}
\end{equation}
where $d\sigma_{eA\rightarrow eX}/dx_B dQ^2$ is the inclusive
DIS cross section. In Eq.~(\ref{rate}), the Bjorken variable
$x_B=Q^2/(2p\cdot q)$ with the virtual photon momentum 
$q^\mu=(k_1-k_2)^\mu$ and $Q^2=-q^2$. 
We work in photon-nucleus frame,  and choose the target momentum 
$p$ along $\vec{z}$-axis, such that 
$p^{\mu}=(p^0,p^x,p^y,p^z)=(P,0,0,P)$, 
and only $p^+=(p^0-p^z)/\sqrt{2}$ is nonvanishing, neglecting target mass.
In this frame, the nucleus is moving in the ``+'' direction.
The struck quark propagates along 
the ``$-$'' direction and could interacts coherently with the 
``remnants'' of the nucleus. The hadron momentum fraction $z_h$ is  
defined as 
\begin{equation}
z_h \equiv \frac{p\cdot p_h }{p\cdot q} = \frac{2x_Bp\cdot p_h}{Q^2} \ .
\label{z}
\end{equation}
With the approximation of one-photon exchange, the semi-inclusive 
DIS cross section 
\begin{equation}
\frac{d{\sigma}_{eA\rightarrow e h X}}{dx_B dQ^2 dz_h}
=\frac{1}{8\pi}\, \frac{e^4}{x_B^2s^2Q^2} \,
L^{\mu\nu}(k_1,k_2)\, \frac{dW_{\mu\nu}}{dz_h}\ ,
\label{sigma-c}
\end{equation}
where $s=(p+k_1)^2$ is the total invariant mass of the lepton-nucleon
system.
In Eq.~(\ref{sigma-c}), the leptonic tensor $L^{\mu\nu}$ is 
given by the diagram in Fig.~\ref{fig1}a,  
\begin{equation}
L^{\mu\nu}(k_1,k_2)
=\frac{1}{2}\, {\rm Tr}(\gamma \cdot k_1 \gamma^{\mu}
\gamma \cdot k_2 \gamma^{\nu}) \ .
\label{L}
\end{equation}

\begin{figure}
\begin{center}
\includegraphics[width=3.0in]{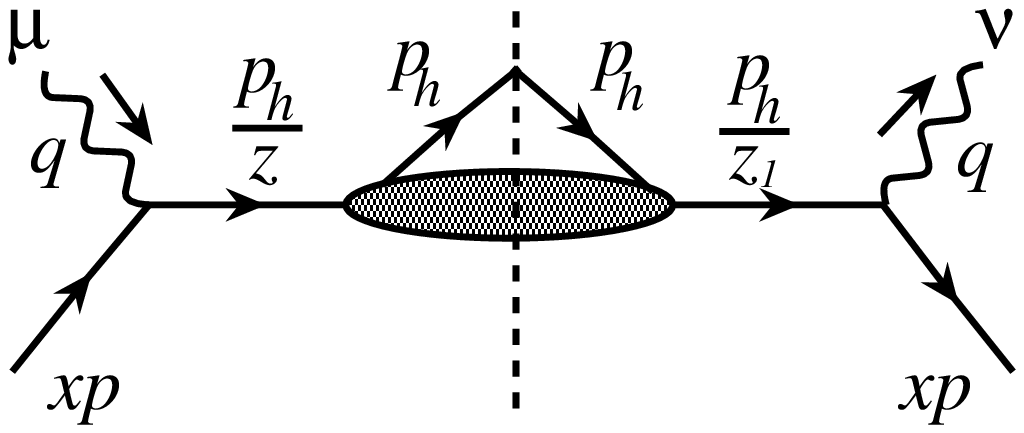}
\hskip 0.5in
\includegraphics[width=2.0in]{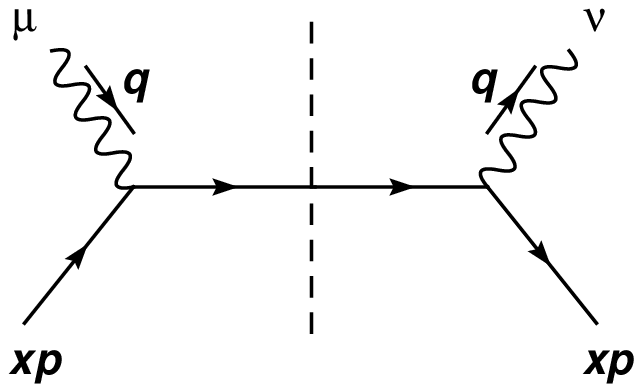}
\caption{Lowest order single scattering Feynman diagram for the
semi-inclusive hadronic tensor with (a) and without (b) the 
nonperturbative parton-to-hadron fragmentation attached.}
\label{fig2}
\end{center}
\end{figure}

The semi-inclusive hadronic tensor $W_{\mu\nu}$ for 
single scattering contribution is represented in Fig.~\ref{fig1}b 
and can be factorized as 
\begin{eqnarray}
\frac{dW_{\mu\nu}}{dz_h}
&=&\sum_q \,\int dz\, dz_1\, 
{\cal D}_{q\rightarrow h}(z,z_1, Q^2) 
\int \frac{dx}{x}\, 
\delta(z-\frac{2xp\cdot p_h}{Q^2})\, 
\delta(z_1-\frac{2xp\cdot p_h}{Q^2})\, 
\nonumber \\
&& {\hskip 0.2in} 
\times \phi_{q}(x,Q^2)\,
H^{(S)}_{\mu\nu}(x,z,z_1)\, ,
\label{dwdz} 
\end{eqnarray} 
where $\sum_q$ runs over (anti)quark flavors, 
$\phi_{q}$ is the leading twist nuclear quark distribution, and 
$H^{(S)}_{\mu\nu}$ is the partonic part. 
In Eq.~(\ref{dwdz}) we introduced a 
two-momentum quark-to-hadron fragmentation density 
${\cal D}_{q\rightarrow h}(z,z_1,Q^2)$, which is defined as   
$\int dz_1\, \delta(z-z_1){\cal D}_{q\rightarrow h}(z,z_1,Q^2) 
= D_{q\rightarrow h}(z,Q^2)$ with $D_{q\rightarrow h}$ 
the normal quark-to-hadron fragmentation function.
The $z_h$-dependence in Eq.~(\ref{dwdz}) is 
implicit in the argument of the $\delta$-function, 
$2x p\cdot p_h/Q^2 = (x/x_B) z_h$.
In Fig.~\ref{fig2} we show the lowest order Feynman diagram 
for the semi-inclusive hadronic tensor on a quark state 
with (a) and without (b) the nonperturbative parton-to-hadron 
fragmentation attached.  Fig.~\ref{fig2}b gives the lowest order 
$H^{(S)}_{\mu\nu}$~\cite{Brock:1993sz}:
\begin{equation}
H^{(S)}_{\mu\nu}=\frac{1}{2}\, 
e^T_{\mu\nu}\,\sum_q\,  x \, e_q^2
\delta \left(x-\frac{Q^2}{2p\cdot q} \right) 
\label{H0}
\end{equation}
where $e_q$ is the  quark fractional charge, and $e^T_{\mu\nu}$ is
defined as 
\begin{equation}
e^T_{\mu\nu}=\frac{1}{p\cdot q}\left[p_\mu q_\nu + q_\mu p_\nu\right]
   +\frac{2x_B}{p\cdot q} p_\mu p_\nu - g_{\mu\nu}\, . 
\label{et-def}
\end{equation} 
Thus, at the lowest order,
\begin{equation}
\frac{dW_{\mu\nu}^{(0)}}{dz_h}=\frac{1}{2}\, e^T_{\mu\nu} \sum_q \,
                        \phi_{q}(x_B,Q^2)\, D_q(z_h, Q^2) \, ,
\label{dwdz0} 
\end{equation} 
with $z_h$ defined in Eq.~(\ref{z}).

\begin{figure}
\begin{center}
\includegraphics[width=5.0in]{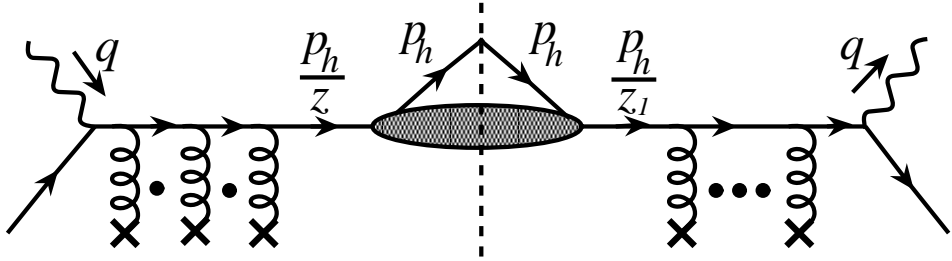}
\caption{Sample diagram shows the interference between two
amplitudes with the same initial and final states but 
different partonic scattering.}
\label{fig_int}
\end{center}
\end{figure}

Single scattering is localized and its medium size dependence is 
limited to that in the nuclear parton distribution function. 
We are interested in additional nuclear effects from 
multiple scattering of the scattered quark and the interference
of amplitudes with different partonic scatterings, as sketched in 
Fig.~\ref{fig_int}. In this section, we first consider the 
double scattering 
contribution to the semi-inclusive hadronic tensor. 

\begin{figure}
\begin{center}
\includegraphics[width=2.5in]{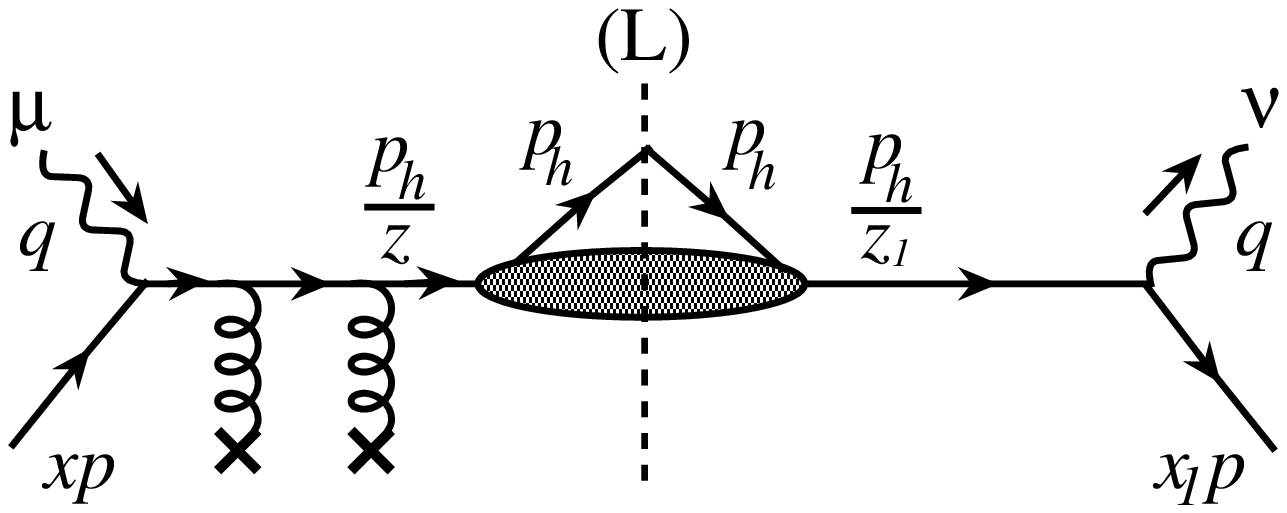}
\hskip 0.5in
\includegraphics[width=2.5in]{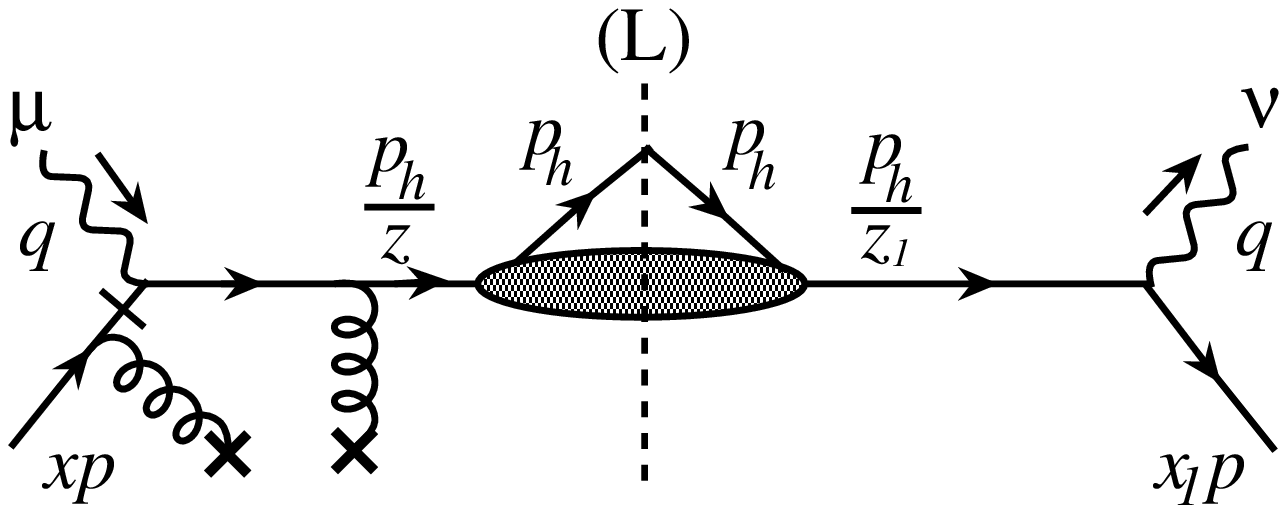}

(a)

\includegraphics[width=2.5in]{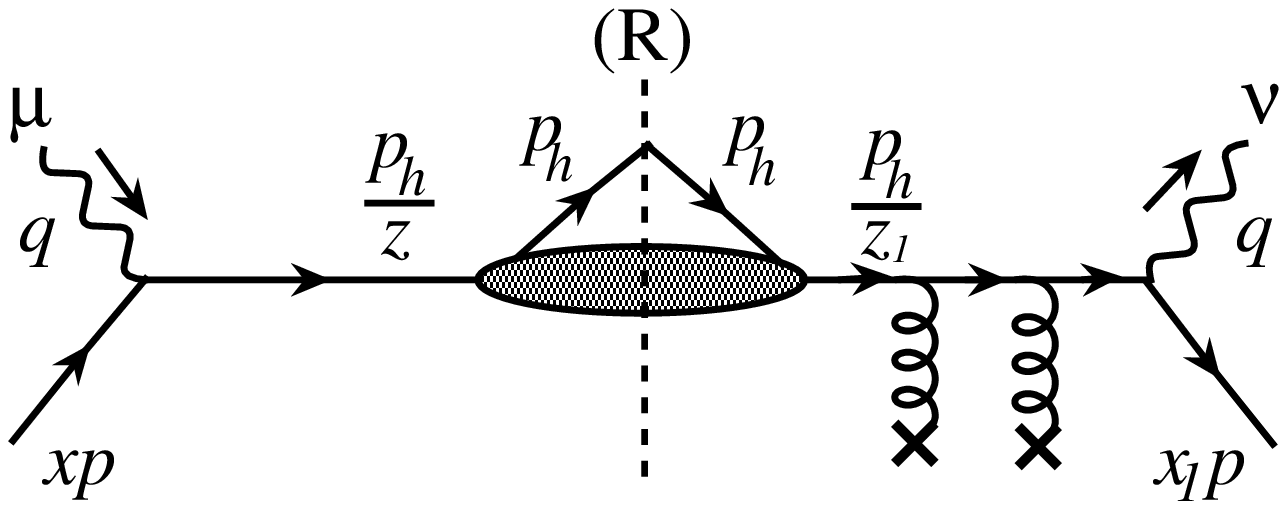}
\hskip 0.5in
\includegraphics[width=2.5in]{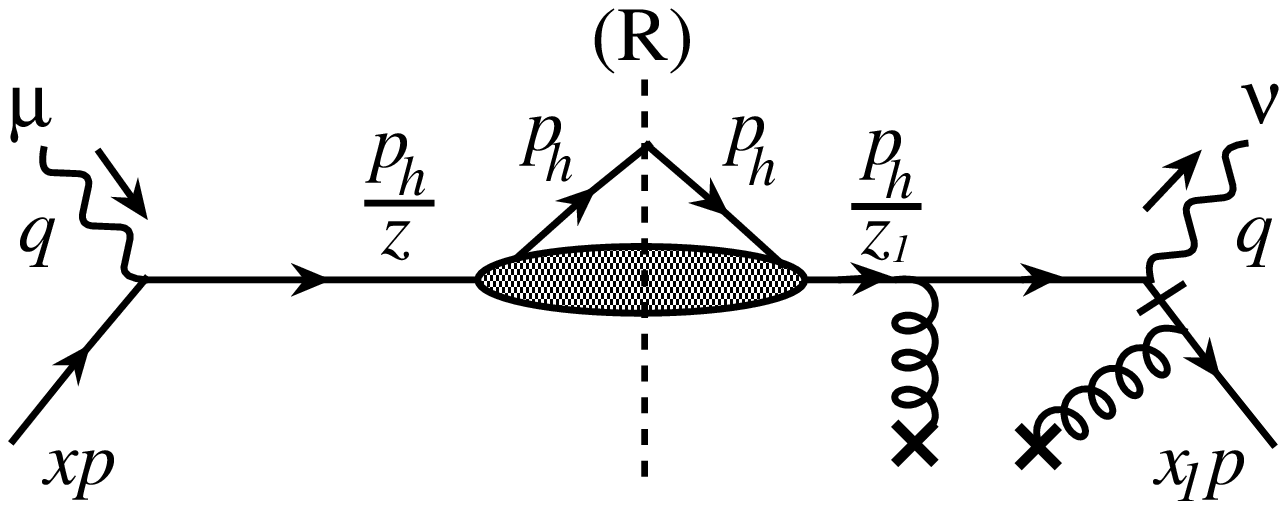}

(b)

\includegraphics[width=2.5in]{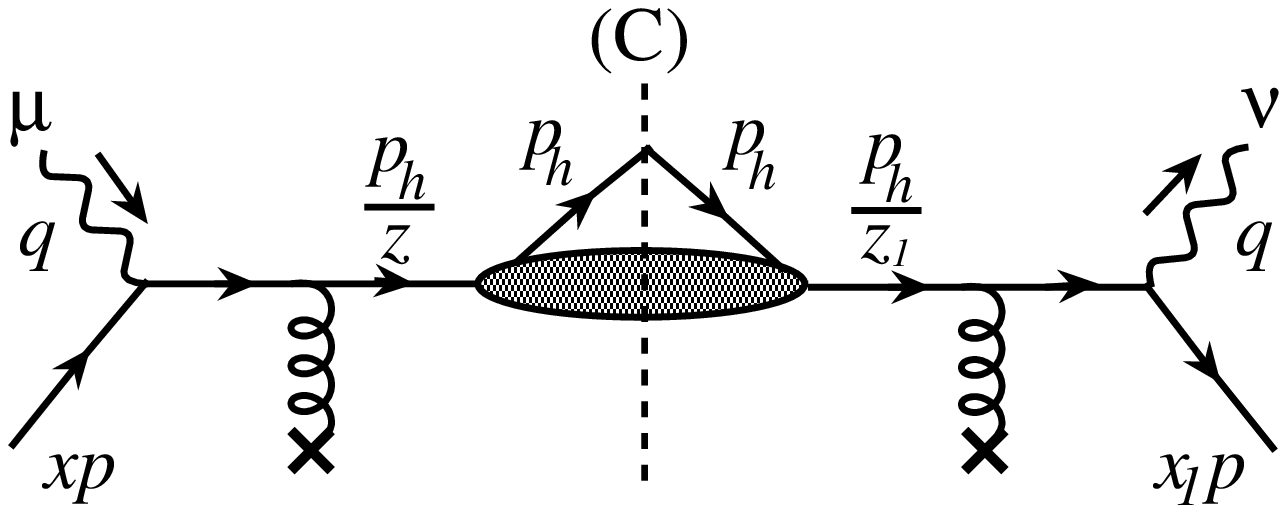}
\hskip 0.5in
\includegraphics[width=2.5in]{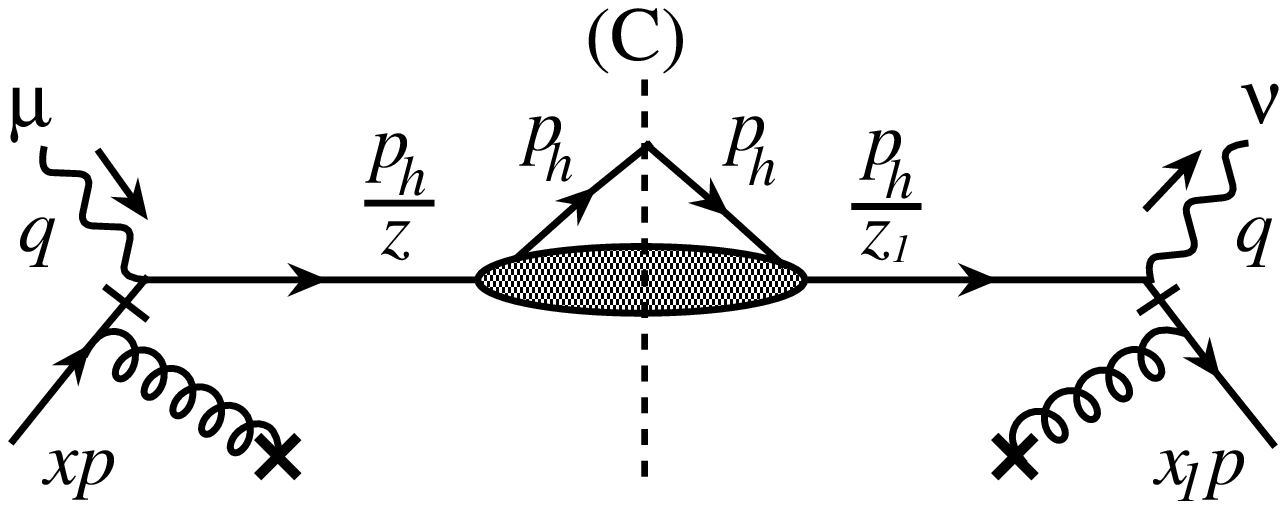}

\includegraphics[width=2.5in]{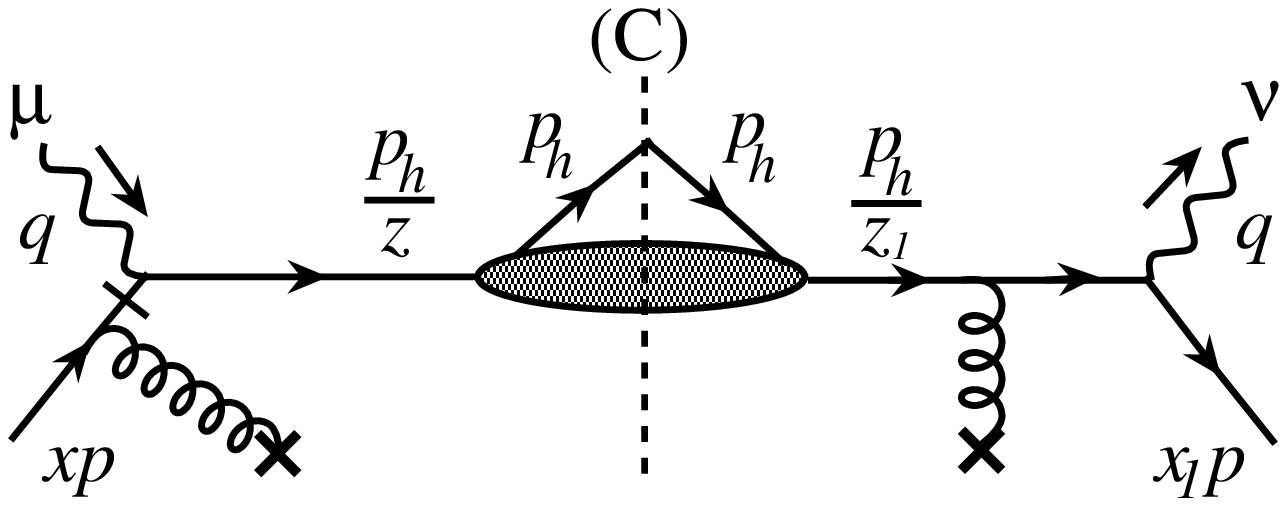}
\hskip 0.5in
\includegraphics[width=2.5in]{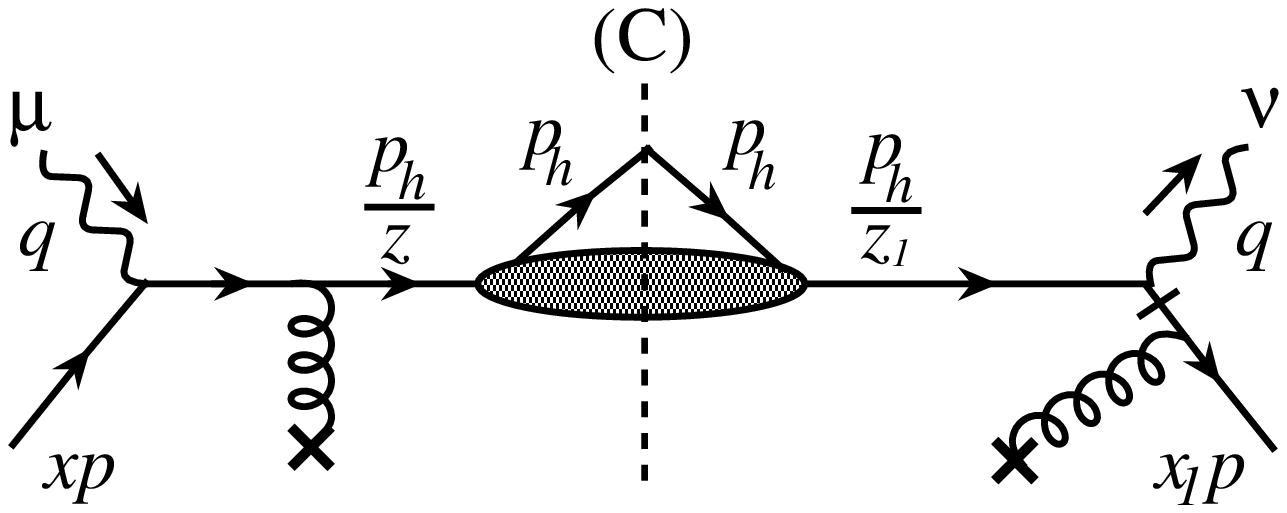}

(c)
\caption{Leading order double scattering Feynman diagrams that
contribute to the semi-inclusive deep inelastic scattering on a 
quark state.  Diagrams of (a) and (b) give explicit $A^{1/3}$-type 
medium size enhancement; and diagrams of (c) give localized 
contributions.}
\label{fig3}
\end{center}
\end{figure}

Fig.~\ref{fig3} shows leading order double scattering 
Feynman diagrams for the semi-inclusive hadronic tensor on 
a quark state. The fermion lines with a short bar represent
the contact terms of quark propagators, and diagrams with both gluons 
attached to the incoming quark line vanish \cite{Qiu}.
Although all diagrams in Fig.~\ref{fig3} could contribute to the 
production rate of the hadron in SIDIS, as we discuss below,
only diagrams in Figs.~\ref{fig3}(a) and (b) can give the leading 
power contribution in the $A^{1/3}$-type medium size 
enhancement \cite{QS-hardprobe}. 
In terms of Feynman diagram, contribution of every propagator 
consists of two parts: a potential pole contribution and a contact 
contribution \cite{Qiu}.  
For example, a quark propagator of momentum $k$ can always be written as  
\begin{equation}
\frac{i\gamma\cdot k}{k^2 + i\epsilon} 
= \frac{i\gamma\cdot \hat{k}}{k^2 + i \epsilon}
+ \frac{i\gamma\cdot n}{2k\cdot n}\, \frac{k^2}{k^2+ i\epsilon}\ ,
\label{DF0}
\end{equation}
where $\hat{k}^2 = 0$ and $n^\mu$ is any auxiliary vector with $k\cdot
n \neq 0$.  The first term in the right-hand-side of Eq.~(\ref{DF0})
corresponds the potential pole contribution when $k^2\rightarrow 0$, 
while the second term is the
contact contribution \cite{Qiu}.  Attaching one gluon to the initial
quark line introduces a quark propagator, and this propagator will
have both the pole and contact contributions.  The pole contribution
is long-distance in nature, representing the interactions between the
quark and the gluon long before the hard collision between the quark
and the virtual photon.  The pole contribution of the incoming quark
propagator should be part of the nuclear quark distribution, and 
is partially responsible for the relatively weak $A$-dependence of the 
leading-twist parton distributions in a nucleus \cite{Qiu:2002mh, CQR}.
On the other hand, the contribution of the contact term is localized 
in space \cite{Qiu}, and does not result into the $A^{1/3}$ type of 
nuclear enhancement \cite{QS_fac}.  Since Feynman diagrams in 
Fig.~\ref{fig3}(c) form a gauge invariant subset of short-distance 
partonic contributions, this set of diagrams will not 
contribute to the leading $A^{1/3}$-type of nuclear enhancement.
Diagrams with the initial-state contact interactions 
in Figs.~\ref{fig3}(a) and (b) vanish, while the other two diagrams 
with final-state rescattering will have at least one pole contribution.

The pole contribution from diagrams with the final-state rescattering, 
shown in Figs.~\ref{fig3}(a) and (b), is responsible for the leading
$A^{1/3}$-type of nuclear enhancement, because taking the residue of 
the unpinched pole effectively sets the gluon momentum to zero and 
leaves the corresponding coordinate space integration of the gluon 
field to the size of nucleus \cite{QS-hardprobe}. Similar to 
Eq.~(\ref{dwdz}), 
the factorized form of the leading pole contribution from the double 
scattering diagrams in Figs.~\ref{fig3}(a) and (b) can be expressed as 
\begin{eqnarray}
\frac{dW_{\mu\nu}}{dz_h} 
&=&
\sum_q \, 
\int dz\, dz_1\,   {\cal D}_{q\rightarrow h}(z,z_1,Q^2)  
\int dx\, d\tilde{x_1}\, dx_1\, 
\delta(z-\frac{2 x p\cdot p_h}{Q^2})\, 
\delta(z_1-\frac{2 x_1 p\cdot p_h}{Q^2}) 
\nonumber \\
&& \times \,
M_A(x,\tilde{x_1},x_1) \,
  H^{(D)}_{\mu\nu}(x,\tilde{x_1},x_1,x_B, z, z_1)\, .
\label{W-qA}
\end{eqnarray}
In Eq.~(\ref{W-qA}), the hadronic matrix element $M_A$ is given by
\cite{QS-hardprobe} 
\begin{eqnarray}
M_A(x,\tilde{x_1},x_1)
&=&
\frac{-1}{x_1-\tilde{x_1}-i\epsilon}\frac{1}{\tilde{x_1}-x-i\epsilon} 
\int \frac{dy^-}{2\pi}\, \frac{dy_1^-}{2\pi}\,
      \frac{d\tilde{y_1}^-}{2\pi}\,
      {\rm e}^{ixp^+y^-}\, {\rm e}^{i(\tilde{x_1}-x)p^+\tilde{y_1}^-}\,
      {\rm e}^{-i(\tilde{x_1}-x_1)p^+y_1^-}
\nonumber \\
&\times &
\langle P_A | \bar{\psi}(0) \frac{\gamma^+}{2}
              F^{+\alpha}(y_1^-)F_{\alpha}^{\ +}(\tilde{y_1}^-)
              \psi(y^-) |P_A\rangle\, .
\label{TqA}
\end{eqnarray}
Here we used $F^{+\alpha}(y^-)=n^\rho \partial_\rho A^\alpha(y^-)$ in 
light-cone gauge. We adapt the $i\epsilon$ prescription introduced in 
Ref.~\cite{MQ-recomb} for the two poles $1/(x_1-\tilde{x_1})$ and 
$1/(\tilde{x_1}-x)$.  
If the partonic part, 
$H^{(D)}_{\mu\nu}(x,\tilde{x_1},x_1,x_B, z, z_1)$, 
is a nonvanishing smooth function at the poles, 
which will be verified later,
the two unpinched poles in Eq.~(\ref{TqA}) can be used to perform 
the contour integration for $d\tilde{x_1}\,dx_1$
in Eq.~(\ref{W-qA}).  In this sense, the hadronic matrix element
$M_A$ can be effectively written as  
\begin{equation}
M_A(x,\tilde{x_1},x_1)
\to
\delta(\tilde{x_1}-x) \,
\delta(x_1-\tilde{x_1}) \, 
{\cal F}_A(x,\tilde{x_1},x_1)
\label{poles}
\end{equation}
with the function ${\cal F}_A$ defined as
\begin{eqnarray}
{\cal F}_A(x,\tilde{x_1},x_1)
&=& 
\int \frac{dy^-}{2\pi}\, \frac{dy_1^-}{2\pi}\,
      \frac{d\tilde{y_1}^-}{2\pi}\,
      {\rm e}^{ixp^+y^-}\, {\rm e}^{i(\tilde{x_1}-x)p^+\tilde{y_1}^-}\,
      {\rm e}^{-i(\tilde{x_1}-x_1)p^+y_1^-}
\nonumber \\
&&\times 
(2\pi)^2 \theta(y_1^{-})\,\theta(\tilde{y_{1}}^{-})\, 
\langle P_A | \bar{\psi}(0) \frac{\gamma^+}{2}
              F^{+\alpha}(y_1^-)F_{\alpha}^{\ +}(\tilde{y_1}^-)
              \psi(y^-) |P_A\rangle\, .
\label{FqA}
\end{eqnarray}

\begin{figure}
\begin{center}
\includegraphics[width=2.5in]{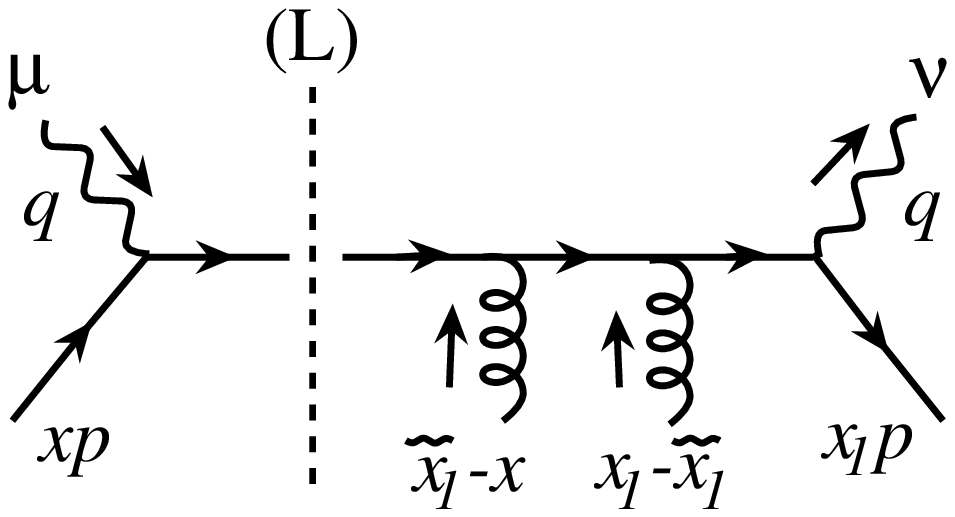}
\hskip 0.5in
\includegraphics[width=2.5in]{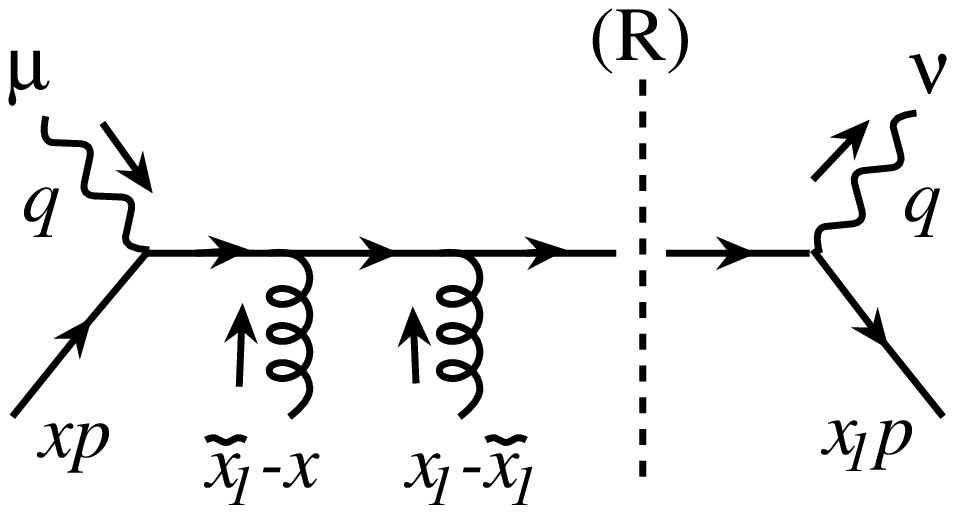}
\caption{Lowest order double scattering Feynman diagrams 
that contribute to the hard part of the 
leading power $A^{1/3}$-type nuclear enhancement.}
\label{fig4}
\end{center}
\end{figure}

The partonic part $H^{(D)}_{\mu\nu}$ in Eq.~(\ref{W-qA}) is given by the 
Feynman diagrams in Fig.~\ref{fig4} with the quark lines contracted by
$\frac{1}{2}\gamma\cdot p$ and gluon lines contracted by $\frac{1}{2}
d_{\alpha\beta}$.  The transverse tensor
$d_{\alpha\beta}=-g_{\alpha\beta} + \bar{n}_\alpha n_\beta + n_\alpha
\bar{n}_\beta$ with two lightlike vectors, $n=(n^+, n^-, n_T)=(0,1,0_{\perp})$
and $\bar{n}=(1,0,0_{\perp})$. 
The leading pole contribution of $H^{(D)}_{\mu\nu}$  with cut 
on the left side is 
\begin{eqnarray}
H^{(D)}_{\mu\nu} \left|_{L-cut}  \right.
&=& e^T_{\mu\nu}\, 
\left(\frac{1}{2}\, e_q^2 \right) \left(\frac{2\pi\alpha_s}{3}\right)
    \frac{1}{Q^2}\, x_B 
      \frac{\delta(x-x_B)}{x_1-x} , 
\label{H-DL}
\end{eqnarray}
and the contribution with cut on the right side is 
\begin{eqnarray}
H^{(D)}_{\mu\nu} \left|_{R-cut} \right. 
&=& e^T_{\mu\nu}\, 
\left(\frac{1}{2}\, e_q^2 \right) \left(\frac{2\pi\alpha_s}{3}\right)
    \frac{1}{Q^2}\, x_B 
      \frac{\delta(x_1-x_B)}{x-x_1} ,
\label{H-DR}
\end{eqnarray}
with the quark fractional charge $e_q$ and 
$e^T_{\mu\nu}$ given in Eq.~(\ref{et-def}).
As we can see, the individual contribution of Eqs.~(\ref{H-DL}) and 
(\ref{H-DR}) is divergent. However, the sum of the contributions 
is finite. 

Summing up contributions from the both cuts, and 
using Eqs.~(\ref{W-qA}), (\ref{poles}),  
(\ref{H-DL}), and (\ref{H-DR}), we have 
the leading order contribution from the double final state scattering:
\begin{eqnarray}
\frac{dW_{\mu\nu}^{(1)}}{dz_h} 
& \propto &
\int dz dz_1\, {\cal D}_{q\to h}(z,z_1, Q^2)  
\int dx\,  dx_1\, 
\delta(z-x\frac{2p\cdot p_h}{Q^2})\,
\delta(z_1-x_1\frac{2p\cdot p_h}{Q^2}) 
\nonumber \\
& & \times 
\delta(x_1-x) {\cal F}_A(x,x,x_1)\, 
\frac{x_B}{Q^2}
\left[
\frac{\delta(x-x_B)}{x_1-x}
+
\frac{\delta(x_1-x_B)}{x-x_1}
\right]
\label{dw-dx}\\
& = &
\int dx\, dx_1\, {\cal F}_A(x,x,x_1)\, 
\int dz\, dz_1\,  
\delta(x-z\frac{Q^2}{2p\cdot p_h})\,
\delta(x_1-z_1\frac{Q^2}{2p\cdot p_h}) 
\nonumber \\
& & \times 
\delta (z-z_1) {\cal D}_{q\to h}(z,z_1,Q^2)\,
\frac{x_B}{Q^2}\, \frac{Q^2}{2p\cdot p_h}
\left[\frac{\delta(z-z_h)}{z_1-z}
 + \frac{\delta(z_1-z_h)}{z-z_1} \right] 
\label{dw-intz}
\end{eqnarray}
with an overall constant factor $(e^{T}_{\mu\nu}/2)\, 
e_q^2\, (2\pi\alpha_s/3)$ from the hard parts in Eqs.~(\ref{H-DL}) 
and (\ref{H-DR}) and a sum over all quark and antiquark flavor.
Because of the $\delta (z-z_1)$, we can expand the $z_1$ in 
above expression inside the square bracket around $z$:
\begin{eqnarray}
\frac{\delta(z-z_h)}{z_1-z}
 + \frac{\delta(z_1-z_h)}{z-z_1}
\approx -\delta'(z-z_h) \ .
\label{dzh}
\end{eqnarray}
Combining Eqs.~(\ref{dw-intz}) and (\ref{dzh}), and carrying out
all integrations by using the $\delta$-functions, we have 
\ba
\frac{dW_{\mu\nu}^{(1)}}{dz_h} &=&  
\frac{1}{2} e_{\mu\nu}^T\, \sum_q e_q^2
\left(\frac{4\pi^2\alpha_s}{3}\right)
    \frac{ x_B}{Q^2}\,\frac{d}{dx_B}T_{qg}^A(x_B,Q^2)\, 
    D_{q\to h}(z_h, Q^2)
\nonumber \\
&+&
\frac{1}{2} e_{\mu\nu}^T \, \sum_q e_q^2
\left(\frac{4\pi^2\alpha_s}{3}\right)
    \frac{ z_h}{Q^2}\,T_{qg}^A(x_B,Q^2)\, 
    \frac{dD_{q\to h}(z_h, Q^2)}{dz_h}
\label{FT-1}
\ea
where $\sum_q$ runs over all quark and antiquark flavors, and the
twist-4 quark-gluon correlation function is defined as 
\cite{LQS2} 
\begin{eqnarray}
T_{qg}^{A}(x_B,Q^2) 
&=&
 \int \frac{dy^{-}}{2\pi}\, e^{ix_Bp^{+}y^{-}}
 \int \frac{dy_1^{-} d\tilde{y}_{1}^-}{2\pi} \,
      \theta(y_1^{-})\,\theta(\tilde{y}_{1}^{-}) 
\nonumber \\
&&\times
\langle P_{A}|\bar{\psi}_{q}(0)\,\frac{\gamma^{+}}{2}\,
    F^{+\alpha}(y_{1}^{-})F_{\alpha}^{\ +}(\tilde{y_{1}}^{-})
    \psi_{q}(y^{-})|P_{A} \rangle \ .
\label{TqF-A}
\end{eqnarray}
This result shows that interference of amplitudes with 
different parton-level multiple scatterings affects the hadron 
production rate in SIDIS.  The effect is sensitive to the
slope of incoming parton flux as well as the shape of the 
fragmentation function.
When combined with the lowest order in Eq.~(\ref{dwdz0}), the
first term in Eq.~(\ref{FT-1}) is responsible for the 
high twist shadowing of inclusive deep inelastic scattering (DIS) 
\cite{Qiu:2003vd}.  The same effect should also appear in the 
denominator of Eq.~(\ref{rate}), and therefore, 
it will not be included in the rest of the discussion for the 
hadron production rate defined in Eq.~(\ref{rate}).
    
From Eq.~(\ref{dw-dx}), we can also derive Eq.~(\ref{FT-1}) 
by first expanding $x_1$ around $x$ utilizing 
the $\delta$-function  $\delta(x_1-x)$:
\begin{equation}
\frac{\delta(x-x_B)}{x_1-x}
 + \frac{\delta(x_1-x_B)}{x-x_1} \approx - \delta' (x-x_B)\, ,
\end{equation}
and then integrating over $dxdx_1$ and $dzdz_1$.
In next section, when we calculate the higher order multiple scattering 
effect, we will follow this derivation to make the presentation simpler.

\section{generalize to higher order multiple scattering}

To compute the effect  of higher order final state multiple scattering,
we add pairs of gluon interactions to the struck quark and  
convert the gluon field operators in the hadronic matrix element of
$W^{\mu\nu}$ to the corresponding field strength.
Each pair of gluon interaction will contribute a factor of~\cite{Qiu:2003vd}
\begin{equation}
   x_B \frac{ 2 \pi \alpha_s}{3Q^2}     
\int \frac{d y_i^-}{2 \pi} \, \frac{d \tilde{y_i}^-}{2 \pi} \,
\frac{e^{i(x_i-\tilde{x_{i}})p^+ y_i^-}}{x_i - \tilde{x_{i}} - i\epsilon} 
\frac{e^{i(\tilde{x_{i}}-x_{i-1})p^+ \tilde{y_i}^-}}
{\tilde{x_{i}}-x_{i-1} - i\epsilon} 
F^{+ \alpha}(y_i^-)  F_{\alpha}^{\; +}(\tilde{y_i}^-)
\left\{
\begin{array}{ll}
\frac{ -1 }{x_{i-1}-x_B +i \epsilon }  & {\rm (L)}   
\\[1ex]
\frac{ -1 }{x_{i}-x_B -i \epsilon }  & {\rm (R)}
\end{array}
\right. . 
\label{ScalVert}
\end{equation}
``L'' (``R'') means the gluon pair are to the left (right) of the final state 
cut line.

To obtain the leading pole contribution for the partonic part with 
$n$ additional 
scattering, we need to sum over all diagrams with all possible 
insertions of the $n$ gluon pairs to both sides of the  
final state cut line. 
Similar to Eq.~(\ref{poles}) we can replace the poles in 
Eq.~(\ref{ScalVert}) by corresponding $\delta$- functions,
expand all $x_i$ of $\delta (x_i-x)$ around $x$, and obtain
\begin{eqnarray}
H^{(n)}_{\mu\nu} 
&=& 
e^T_{\mu\nu}\, 
\left(\frac{1}{2}\, e_q^2 \right) 
\left(\frac{2\pi\alpha_s}{3}\right)^n
    \left(\frac{x_B}{Q^2}\right)^n \, (-1)^n
\frac{d^n}{dx}\delta (x-x_B)\, . 
\label{H-n}
\end{eqnarray}
Convoluting $H^{(n)}_{\mu\nu}$ with the fragmentation function and 
following similar derivations for the double scattering case, 
we obtain the leading pole contribution for semi-inclusive hadronic 
tensor, with $n$ additional scattering:
\begin{equation}
\frac{ dW_{\mu\nu}^{(n)}}{dz_h} 
\approx  
\frac{1}{2} e^T_{\mu\nu}
\sum_q e_q^2 \left[ z_h \frac{4\pi^2 \alpha_s}{3 Q^2} \right]^n  
\! 
\frac{1}{n!}  \, M_A^{(n)}(x_B, Q^2)\, 
\frac{d^n}{dz_h^n} D_{q\to h}(z_h, Q^2)\,
\label{FTmel}   
\end{equation}
with the multi-field matrix element $M_A^{n}$ given by 
\begin{equation}
M_A^{n}(x,Q^2) = 
\int \frac{d y_0^- }{2\pi}\, e^{i x p^+ y_0^-} \, 
\langle P_A |\bar{\psi}_f(0)\,  \frac{\gamma^+}{2}\,  
 \psi_f(y_0^-) \,   
 \prod\limits_{i=1}^{n} 
\left[  \int p^+d y_i^-  \, \theta(y_i^-) \hat{F}^2(y_i^-)  \right] 
| P_A \rangle \, .  
\label{Mn}
\end{equation}
The integration $\int p^+dy_i^-$ in Eq.~(\ref{Mn}) gives the nuclear
size dependence~\cite{Qiu:2002mh}. And the operator $\hat{F}^2(y_i^-)$  
is given by 
\begin{equation}
\hat{F}^2 (y_i^-) \equiv  
\int \frac{d  \tilde{y_i}^-}{2 \pi} \; 
\frac{F^{+ \alpha}(y_i^-)  F_{\alpha}^{\; +}(\tilde{y_i}^-)}{p^+} 
\, \theta(\tilde{y_i}^-) \; .
\label{FFlambda}
\end{equation}
Compare with the operator definition of gluon density, we can see
that its expectation value 
can be related to the small-$x$ limit of the gluon distribution,
$\langle p | \hat{F}^2 (y_i^-) | p \rangle  
\approx \lim_{x \rightarrow 0} \, \frac{1}{2} \, x \, G(x,Q^2)$,
and is independent of $y_i$ \cite{Qiu:2003vd}.
 
In order to evaluate the multi-field matrix element 
in Eq.~(\ref{Mn}), we approximate the expectation value of 
the product of operators to be a product of expectation
values of the basic operator units in a nucleon state of momentum 
$p = P_A/A$:        
\begin{equation}
\langle P_A | \, \hat{O}_0  \, \prod\limits_{i=1}^{n}  
 \hat{O}_i  \,  | P_A \rangle =
A \,\langle p \, | \, \hat{O}_0 \, |  \, p \rangle 
\prod\limits_{i=1}^{n} \left[ N_p \,  
\langle p \, | \, \hat{O}_i \, | \,  p \rangle \right] \;, 
\nonumber 
\end{equation}
where $N_p$ is the normalization. In a model of constant lab frame nucleus 
density $\rho(r)= 3/(4 \pi r_0^3)$, we have 
\be
\int p^+dy_i^- \theta(y_i^-) N_p  
\langle p | \hat{F}^2 (y_i^-) | p \rangle  
=  \frac{9}{16\pi r_0^2} (A^{1/3}-1) 
\langle p | \hat{F}^2 (y_i^-) | p \rangle  
\label{lambda2}
\ee
The factor $(A^{1/3}-1)$ is taken such that the nuclear effect 
vanishes for $A=1$.  With the above model for $M_A^{(n)}$, we have
\be
\frac{ dW_{\mu\nu}^{(n)}}{dz_h} 
\approx 
\frac{1}{2} e^T_{\mu\nu} 
\sum_q e_q^2\, A\, \phi_{q}(x_B,Q^2)\,
\left[  \frac{z_h\, \kappa^2 (A^{1/3}-1)}{Q^2} \right]^n  
\! 
\frac{1}{n!}  \,\frac{d^n}{dz_h^n} D_{q\to h}(z_h, Q^2)\,
\label{W-n}   
\ee
The quantity $\kappa^2$ represent the characteristic scale of quark 
interaction with the medium \cite{Qiu:2003vd}
\begin{eqnarray}
\kappa^2  & = & \frac{3 \pi  \alpha_s (Q^2)}{4\, r_0^2} 
\langle p| \, \hat{F}^2 (y_i) \,| p \rangle  \; .
\label{kappa2}
\end{eqnarray}
Summing the $A^{1/3}$-enhanced contributions in Eq.~(\ref{W-n}) to 
all order in $n$, we have
\begin{eqnarray} 
\frac{ dW_{\mu\nu}}{dz_h}  \! & \approx & \! 
\frac{1}{2} e^T_{\mu\nu}\sum_q e_q^2\, A\, \phi_{q}(x,Q^2)\,
\sum_{n=0}^{N}  \frac{1}{n!}
\left[ \frac{z_h\, \kappa^2 (A^{1/3}-1)}{Q^2} \right]^n  \, 
\frac{d^{n}  D_{q\to h}(z_h,Q^2) }{d^n z_h}   
\nonumber  \\
\! &\approx & \! 
A \, \frac{1}{2}\, e^T_{\mu\nu} \sum_q e_q^2\, \phi_{q}(x,Q^2)\,
D_{q\to h}
\left( z_h+ \frac{z_h\,\kappa^2 ( A^{1/3}-1) }{Q^2}, Q^2 \right) \, ,
\label{W-res}  
\end{eqnarray}
where  $N$ is the upper limit on the number of quark-nucleon interactions.
In deriving Eqs.~(\ref{W-res}) we have taken
$N\approx \infty$ because the effective value of $\kappa^2$ is
relatively small.
Eq.~(\ref{W-res}) is the main result of this paper. 
It shows that the net effect of multiple scattering without induced 
radiation for a propagating quark in the medium 
is equivalent to a shift in the variable $z$ for the quark 
fragmentation function $D_{q\to h}(z)$, which leads to a
suppression of the hadron production rate.  This is a result of the
quantum interference of amplitudes of multiple scattering.
Such a shift in $z$ for the fragmentation  
function is very similar to the effect of the parton energy loss 
model proposed in Ref.~\cite{Wang:1996}.  The shift 
\be
\Delta z(z_h)= z_h\, \frac{\kappa^2 ( A^{1/3}-1) }{Q^2}
\label{dz}
\ee
depends on only one parameter $\kappa^2$ and the medium length.
The parameter $\kappa^2 \propto \lim_{x\rightarrow0} x G(x,Q^2)$
with $G(x,Q^2)$ the gluon distribution function. 
It can be related to the $\lambda^2$ in LQS model for twist-4 quark-gluon
correlation function $T_{qg}(x)=A^{4/3}\lambda^2 \phi_q(x)$ \cite{LQS2} 
by $\kappa^2=(4\pi^2\alpha_s/3)\lambda^2$. The $\lambda^2$ has been 
estimated using Drell-Yan transverse momentum broadening and DIS 
momentum imbalance \cite{LQS2, Guo1}, and was in the range 
of $0.01-0.1$ GeV$^2$.

Using our result given in Eq.~(\ref{W-res}) ,we obtain the
hadron production rate defined in Eq.~(\ref{rate}) for SIDIS on
a nucleus target $A$:
\begin{eqnarray}
R^A & \approx & 
\frac{\sum_q e_q^2 \phi_q^A(x_B, Q^2)\, 
             D_{q\to h}(z_h+\Delta z(z_h))}
     {\sum_q e_q^2 \phi_q^A(x_B, Q^2)} \ ,
\label{R-A}
\end{eqnarray}
with $\Delta z(z_h)$ given in Eq.~(\ref{dz}).

\section{The quark mass effect}

\begin{figure}
\begin{center}
\includegraphics[width=3.5in]{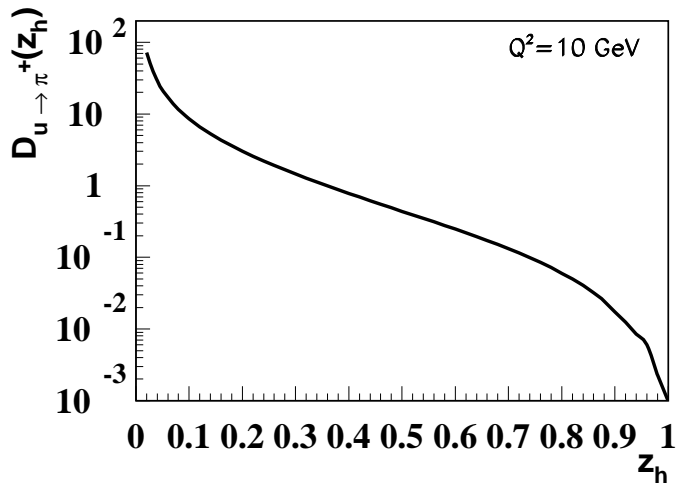}
\caption{ $u$-quark fragmentation function given by 
Ref.~\protect{\cite{Frag}}.}
\label{fignum7}
\end{center}
\end{figure}

\begin{figure}
\begin{center}
\includegraphics[width=3.5in]{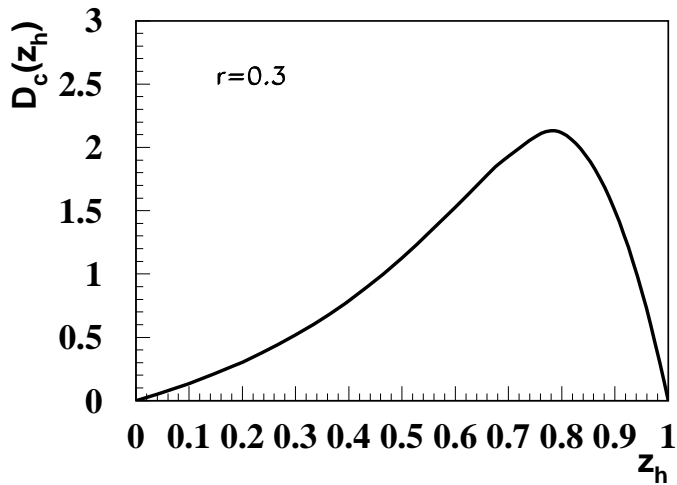}
\caption{Charm quark fragmentation function 
at $Q^2=9$ GeV$^2$ from Ref.~\protect{\cite{Braaten}}.
$r$ is a parameter that represents the ratio of the constituent 
mass of the light quark to the meson mass. }
\label{fignum6}
\end{center}
\end{figure}

In the above derivation, we concentrated on the multiple scattering 
of light quarks and ignored the quark mass. 
If the stuck quark is a heavy quark, we then can not ignore the 
quark mass. In this case, the initial quark momentum can not
be approximated as $xp$. Instead, we have the quark momentum
\begin{equation}
k^{\mu}=xp^+\bar{n}^{\mu}+\frac{m^2}{2xp^+}n^{\mu} \ ,
\label{km}
\end{equation}
with $m$ the quark mass. 
Due to the on-shell condition $(k+q)^2=m^2$, the final state 
$\delta$-function is modified as 
\begin{equation}
\delta((k+q)^2-m^2)= 
\frac{1}{\sqrt{1+4m^2/Q^2}}\,
\frac{x_{Bm}}{Q^2}\, \delta(x-x_B)
\end{equation}
with
\begin{equation}
x_{Bm}=x_B\,\frac{1+\sqrt{1+4m^2/Q^2}}{2} \ .
\label{xm}
\end{equation}
In addition, for additional scattering with each gluon pair in the medium, 
the interaction will contribute a factor  
\begin{eqnarray}
&& 
x_{Bm}\,  
\left[\frac{2}{1+\sqrt{1+4m^2/Q^2}}\right]^2
\frac{ 2 \pi \alpha_s}{3Q^2}
\label{mfactor} \\     
&& {\hskip 0.2in}
\times
\int \frac{d y_i^-}{2 \pi} \, \frac{d \tilde{y_i}^-}{2 \pi} \,
\frac{e^{i(x_i-\tilde{x_{i}})p^+ y_i^-}}{x_i - \tilde{x_{i}} - i\epsilon} 
\frac{e^{i(\tilde{x_{i}}-x_{i-1})p^+ \tilde{y_i}^-}}
{\tilde{x_{i}}-x_{i-1} - i\epsilon} 
F^{+ \alpha}(y_i^-)  F_{\alpha}^{\; +}(\tilde{y_i}^-)
\left\{
\begin{array}{ll}
\frac{ -1 }{x_{i-1}-x_{Bm} +i \epsilon }  & {\rm (L)}   
\\[1ex]
\frac{ -1 }{x_{i}-x_{Bm} -i \epsilon }  & {\rm (R)}
\end{array}
\right. . 
\nonumber
\end{eqnarray}
Correspondingly, due to the multiple scattering with the medium, 
the shift $\Delta z(z_h)$ in fragmentation function 
for a heavy quark with mass $m$ is 
\be
\Delta z_m(z_h) = z_h\, 
\left[\frac{2}{1+\sqrt{1+4m^2/Q^2}}\right]^2
\frac{\kappa^2 ( A^{1/3}-1) }{Q^2} \ .
\label{dzm}
\ee
From Eq.~(\ref{dzm}), we see that the $z$-shift for a heavy quark 
has a similar functional form to that of a light quark, except that 
it has an extra factor that depends on the quark mass. This mass
dependent factor makes the $z$-shift smaller for a heavier quark.  
However, as we can see from Fig.\ref{fignum7} and Fig.\ref{fignum6},
the fragmentation function of a heavy quark have very 
different $z$-dependence from that of a light quark.  
The $z$-shift in fragmentation function can result in very different 
hadron production ratio, even though they have similar forms of 
 $z$-shift.  Therefore, we expect that the net effect of multiple 
scattering for heavy meson production to be very different from that 
for light mesons. 

\section{Numerical estimates and comparision with data} 

In order to compare with data \cite{hermes}, we compute the ratio 
of $R^A$ for a nuclear target $A$ to that of a deuterium target $D$:
\ba
R_M &=& \frac{R^A}{R^D} \nonumber \\
& \approx & 
\frac{\sum_q e_q^2\, \phi_q^A(x_B, Q^2)\, D_{q\to h}(z_h+\Delta z(z_h))}
     {\sum_q e_q^2\, \phi_q^D(x_B, Q^2)\, D_{q\to h}(z)} \,
\frac{\sum_q e_q^2\, \phi_q^D(x_B, Q^2)}
     {\sum_q e_q^2\, \phi_q^A(x_B, Q^2)} \ .
\label{RR}
\ea 
The super script ``A'' and ``D'' represent the nuclear target of 
atomic weight $A$ and the deuterium target, 
respectively.

\begin{figure}
\begin{center}
\includegraphics[width=4.0in]{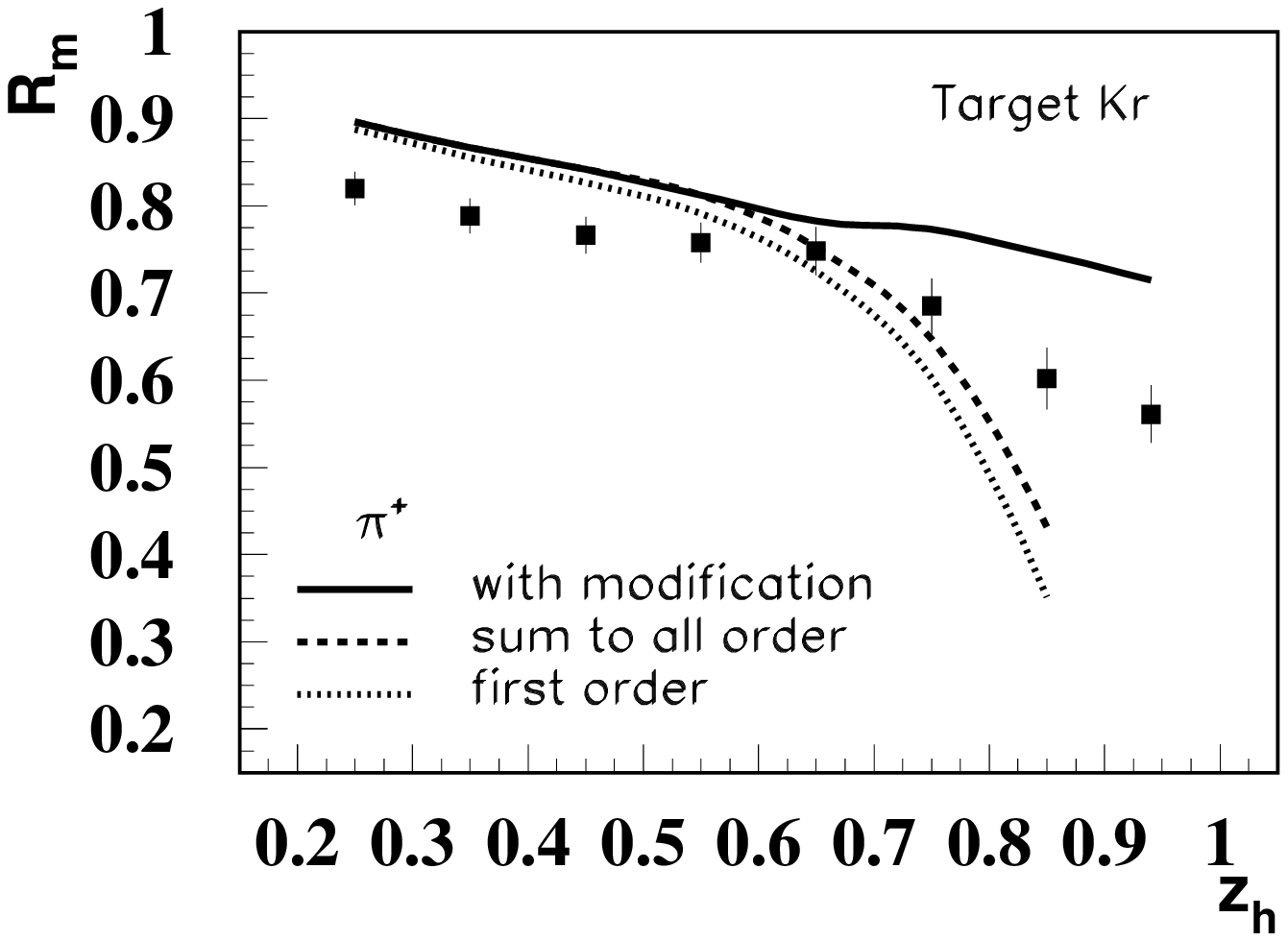}
\caption{The double ratio $R_M$ of Eq.~(\protect{\ref{RR}}) 
for $\pi^+$ production with Krypton
target, compared with HERMES data.}
\label{fignum1}
\end{center}
\end{figure}

\begin{figure}
\begin{center}
\includegraphics[width=4.0in]{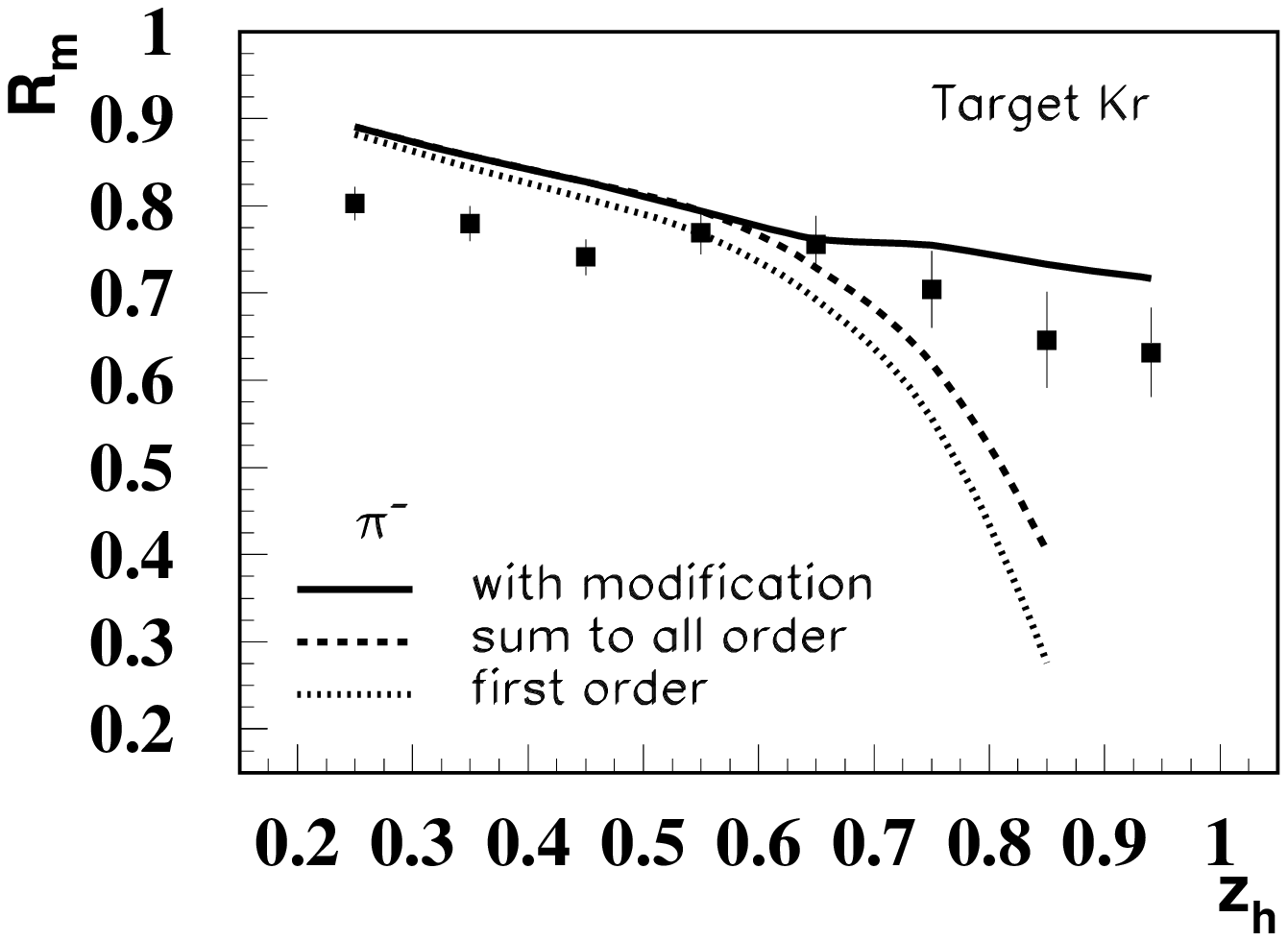}
\caption{The double ratio $R_M$ of Eq.~(\protect{\ref{RR}}) 
for $\pi^-$ production with Krypton
target, compared with HERMES data.}
\label{fignum2}
\end{center}
\end{figure}

\begin{figure}
\begin{center}
\includegraphics[width=4.0in]{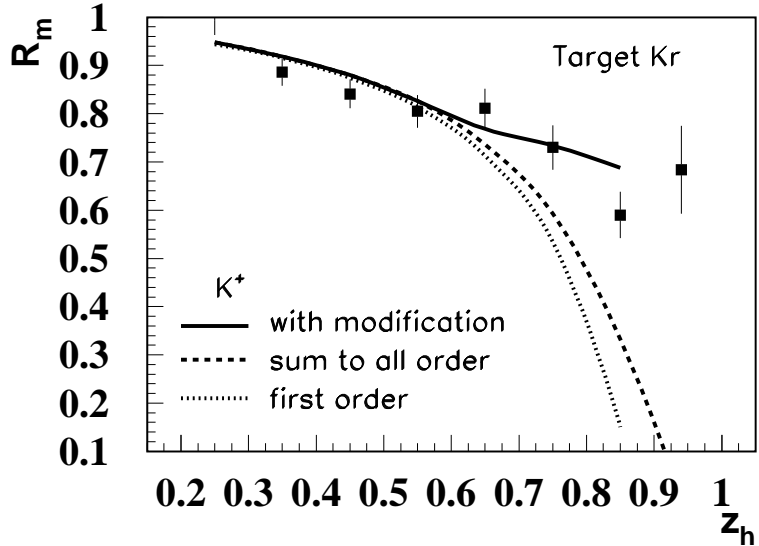}
\caption{The double ratio $R_M$ of Eq.~(\protect{\ref{RR}}) 
for $K^+$ production with Krypton
target, compared with HERMES data.}
\label{fignum3}
\end{center}
\end{figure}

\begin{figure}
\begin{center}
\includegraphics[width=4.0in]{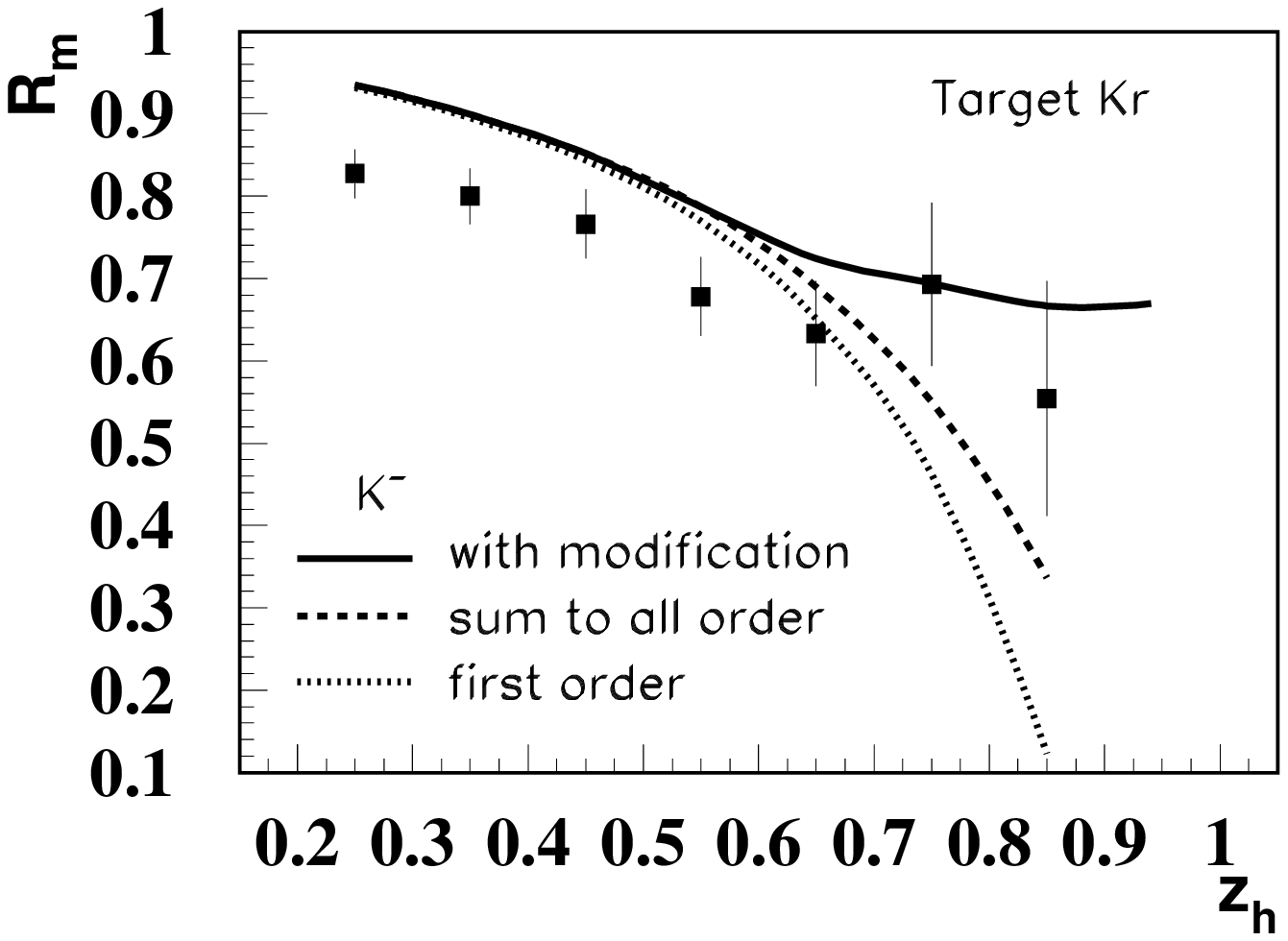}
\caption{The double ratio $R_M$ of Eq.~(\protect{\ref{RR}})  
for $K^-$ production with Krypton
target, compared with HERMES data.}
\label{fignum4}
\end{center}
\end{figure}

To obtain the numerical estimate of the double ratio $R_M$ in
Eq.~(\ref{RR}), we use the lowest order 
CTEQ6 parton distributions \cite{CTEQ}. For the nuclear dependence of 
parton distribution, we use the parameterizations given in 
Ref.~\cite{Eskola}.   
Figs.~\ref{fignum1}-\ref{fignum4} compares our result with the data 
from HERMES experiment\cite{hermes}. The experiment data points have 
$x_B$ in the range of $0.084-0.1$, and the $Q^2\sim 2.2-2.6$ GeV$^2$.
In plotting Figs.~\ref{fignum1}-\ref{fignum4}, we used the 
fragmentation function provided by Ref.~\cite{Frag}.  
The dot lines represent the
result with double scattering only. The dashed curves represent the result when
we sum to all order. We can see that our curve is slightly 
above the data, because our calculation only include  multiple scattering
without induced radiation. The induced radiation will give further 
suppression \cite{Guo:2000nz, Wang:2001if} and brings down the curve. 
We also notice that at large $z_h$ region, our curve is steeper than the 
data point. This is because our result sums the additional scattering to all 
order and it is not applicable for large $z_h$. At large $z_h$, the hadron 
forms early, and have shorter formation time. In this case,  
Summing the additional scatterings to all order is an unrealistic 
approximation. In order to take into account of the hadron formation time, 
which is proportional to $(1-z_h)$ \cite{alberto}, we modify the
$z_h$ shift to be  at large $z_h$
\be
\Delta z (z_h)= z_h \,
\frac{\kappa^2 ( A^{1/3}-1) }{Q^2}\, 
\frac{(1-z_h)}{1-z_c} 
\hspace{0.5cm} {\rm when} 
\hspace{0.5cm} z > z_c \ . 
\label{dz-m}
\ee
In Eq.~(\ref{dz-m}), $z_c$ is a parameter. In this model, we assume that
when $z_h<z_c$, the hadrons will form outside of the nucleus, and we do not 
need to worry about the hadron formation time. We can apply our resumed result
of Eq.~(\ref{dz}) when $z_h<z_c$. When $z_h>z_c$, the hadron formation 
process may start early, and we use the modified $\Delta z(z_h)$ given in 
Eq.~(\ref{dz-m}) to take into account of the 
formation time. The solid lines in Figs.~\ref{fignum1}-\ref{fignum4} 
show our estimates when we use the above 
modified $\Delta z$, and choose $z_c=0.6$. 
The curve is above the data points at large $z_h$, because
here we did not consider the nuclear absorption of the pre-hadron state.  
The nuclear absorption of pre-hadron state should give 
additional suppression and bring down the curve a little more
\cite{alberto}.

\begin{figure}
\begin{center}
\includegraphics[width=4.0in]{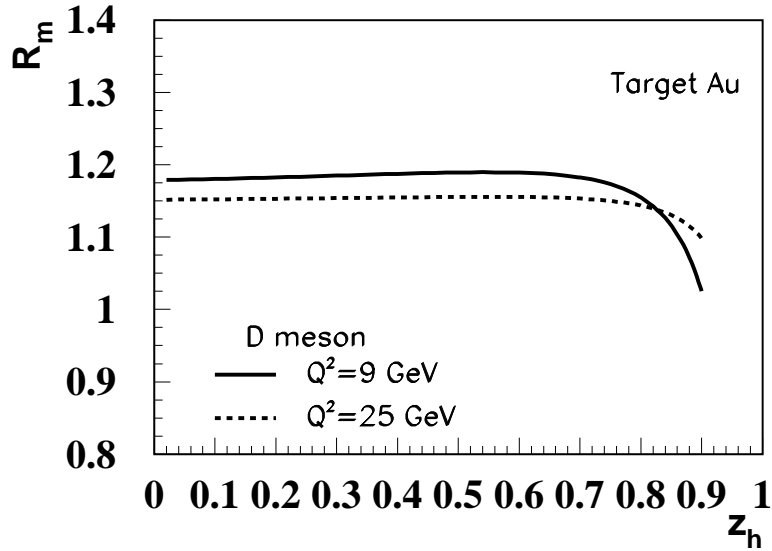}
\caption{Our estimate of  double ratio $R_M$ 
for D meson production with gold
target at $Q^2=9$ GeV$^2$ and $Q^2=25$ GeV$^2$.}
\label{fignum5}
\end{center}
\end{figure}

In Fig.~\ref{fignum5}, we plot the figure for D meson production in 
eA scattering. Due to the limit in collision energy,  there is no data for 
D meson production from  Hermes experiment. However, future 
Electron-ion Collider (EIC) experiments at Brookhaven should be able to 
observe the semi-inclusive D meson production \cite{EIC}. In obtaining 
Fig.~\ref{fignum5},    
we used the $c$-quark fragmentation function of Ref.~\cite{Braaten}. 
From Eq.(39), we see that the ratio $R_M$ is mainly determined by the shift
$\Delta z(z_h)$ and the shape of fragmentation functions.  Since the
fragmentation function for light- and heavy-meson have very different
$z$-dependence, as shown in Fig.\ref{fignum7} and Fig.\ref{fignum6}, 
the shift in $z_h$ results in very different  double ratio $R_M$. 
At smaller $z_h$ region, $z_h$ may actually result in $R_M > 1$ for 
D mesons, due to the characteristic
shape of the $c$-quark fragmentation function, as we can see from 
Fig.~\ref{fignum6}.  

In Fig.~{\ref{fignum1}-\ref{fignum5}, we only illustrate the size of the 
suppression due to the shift of $z_h$, which is caused by coherent multiple 
scattering without induced radiation. A more complete analysis
should include other effects, such as induced radiation.  
\section{Summary and Outlook}

In summary, we demonstrated that 
the coherent multiple scattering of a propagating quark 
with the medium, without the induced radiation, 
can also change the quark fragmentation or 
hadron's production rate.  The net effect of 
leading power contributions in medium length 
is equivalent to a shift in the fragmentation 
function's $z_h \rightarrow z_h + \Delta z$.
At the lowest order, the shift $\Delta z$ is given by 
an universal matrix element, which is proportional to 
$\lim_{x\rightarrow0} x G(x,Q^2)$.   We also show
that for a quark with mass m, the shift will be smaller. 
Our result could be interpreted as the 
collisional energy loss, which is complementary to the 
energy loss of induced radiation.  
However, beyond the leading order, the separation 
of the collisional energy loss and that of induced 
radiation will depend on the factorization scheme and 
will not be unique, and need further study.

In this paper, we derived the effect of the radiationless 
multiple scatterring in semi-inclusive DIS, our approach can be 
systematically generalized to hadron production in $p+A$ and $A+A$ 
collisions.  Because of the convolution of two parton distributions, 
which are steep falling functions of parton momentum fraction $x$, 
the hadron production in 
hadronic collisions is dominated by the large $z$ part of fragmentation 
functions, in particular, for heavy meson production \cite{Berger:2001wr}.  
As shown in Fig.~\ref{fignum6}, the heavy quark fragmentation 
functions are steep falling
functions of $z$ in large $z$ region.  As a result, depending on the 
momentum of the observed heavy meson, or the effective range of $z$, we 
expect that the radiationless multiple scattering will lead to a 
suppression of heavy meson production, similar to that in light 
hadron production.

\section*{Acknowledgments}

This work was support in part by the U.S. National Science Foundation
under Grant PHY-0340729.


\end{document}